\documentclass[twocolumn,10pt,floatfix,longbibliography,letterpaper]{article}

\usepackage[letterpaper,margin=1.5cm,footskip=0.85cm]{geometry}
\usepackage{graphicx}
\usepackage{amsmath,amssymb}
\usepackage{hyperref}
\usepackage{caption}
\usepackage{subcaption}
\usepackage{siunitx}
\usepackage[hang,flushmargin]{footmisc}
\usepackage{authblk}
\usepackage{titlesec}
\titleformat{\section}[hang]
  {\normalfont\large\bfseries}{\thesection}{0.7em}{\setlength{\parindent}{0pt}}
\titleformat{\subsection}
  {\normalfont\normalsize\bfseries}{\thesubsection}{0.75em}{}
\titleformat{\subsubsection}
  {\normalfont\normalsize\itshape}{\thesubsubsection}{0.75em}{}
\titlespacing*{\section}{0pt}{4ex plus .5ex}{1ex}
\titlespacing*{\subsection}{0pt}{1.5ex plus .5ex}{0.75ex}

\newcommand{\blocktitle}[1]{%
  \vspace{0.5\baselineskip}%
  \noindent\textbf{#1}%
}

\usepackage[font=small,labelfont=sc,labelsep=period]{caption}

\usepackage{etoolbox}

\makeatletter
\patchcmd{\@startsection}
  {\@afterindenttrue}
  {\@afterindentfalse}
  {}{}
\makeatother

\apptocmd{\thebibliography}{%
  \setlength{\itemsep}{0pt}%
  \setlength{\parskip}{1pt}%
}{}{}

\let\oldthebibliography\thebibliography
\renewcommand{\thebibliography}[1]{%
  \oldthebibliography{#1}%
  \small
}

\title{An integrated quantitative single-objective light-sheet microscope\\ for subcellular dynamics in embryos and cultured multicellular systems}

\author[1]{Armin Shoushtarizadeh$^*$}
\author[1]{Michele Cerminara$^{*,\dagger}$}
\author[1]{Corinne Chureau}
\author[1]{Leah Friedman}
\author[2]{Deepthi Kailash}
\author[1,2]{Thomas Gregor$^{\dagger}$}

\affil[1]{Department of Developmental and Stem Cell Biology,\protect\\
CNRS UMR3738 Paris Cité, Institut Pasteur, 75015 Paris, France}
\affil[2]{Joseph Henry Laboratories of Physics \& Lewis-Sigler Institute for Integrative Genomics,\protect\\ Princeton University, Princeton 08544, USA}

\date{\today}

\begin{document}

\twocolumn[
\begin{@twocolumnfalse}

\maketitle

\begingroup
\renewcommand\thefootnote{}
\setlength{\footnotemargin}{0pt}
\footnotetext{$^{\dagger}$These authors contributed equally to this work.}
\footnotetext{$^{\dagger}$Corresp.: michele.cerminara@pasteur.fr, thomas.gregor@pasteur.fr}
\endgroup

\begin{center}
\begin{minipage}{0.85\linewidth}

\begin{abstract}
Quantitative imaging of subcellular processes in living embryos, stem-cell systems,
and organoid models requires microscopy platforms that combine high spatial resolution,
fast volumetric acquisition, long-term stability, and minimal phototoxicity.
Single-objective light-sheet approaches based on oblique plane microscopy (OPM) are
well suited for live imaging in standard sample geometries, but most existing
implementations lack the optical calibration, timing precision, and end-to-end
integration required for reproducible quantitative measurements.

Here we present a fully integrated and quantitatively characterized OPM platform
engineered for dynamic studies of transcription and nuclear organization in embryos,
embryonic stem cells, and three-dimensional culture systems. The system combines
high–numerical-aperture remote refocusing with tilt-invariant light-sheet scanning
and hardware-timed synchronization of laser excitation, galvo scanning, and camera
readout. We provide a comprehensive characterization of the optical performance,
including point-spread function, sampling geometry, usable field of view, and system
stability, establishing a well-defined framework for quantitative volumetric imaging.

To support high-throughput operation, we developed a unified acquisition and
reconstruction pipeline that enables real-time volumetric imaging at hardware-limited
rates while preserving deterministic timing and reproducible geometry. Using this
platform, we demonstrate quantitative three-dimensional imaging of MS2-labeled
transcription sites in living \textit{Drosophila} embryos, cultured mouse embryonic
stem cells, and mESC-derived gastruloids, enabling extraction of transcriptional
intensity traces across diverse biological contexts.

Together, this work establishes OPM as a robust and quantitatively calibrated
single-objective light-sheet platform for transcription imaging in complex living
systems, providing a foundation for future studies of gene regulation and nuclear
dynamics across developmental and stem-cell models.
\end{abstract}

\end{minipage}
\end{center}

\vspace{1cm}

\end{@twocolumnfalse}
]


\section*{Introduction}
\label{sec:intro}

Quantitatively imaging subcellular processes in living systems with high spatial and temporal resolution is central to modern developmental and cellular biology. Many of the most dynamic regulatory events—transcriptional bursting, enhancer–promoter communication, chromatin motion, mRNA synthesis and transport, and mitotic reorganization of nuclear structure—unfold over seconds to minutes and within crowded three-dimensional environments.  Capturing these processes in intact embryos, tissues, and stem-cell–derived organoids requires imaging modalities that combine high spatial resolution, fast volumetric acquisition, low photodamage, and long-term stability under physiological and cell-culture-friendly conditions. Meeting these requirements remains a major experimental challenge.

Light-sheet fluorescence microscopy (LSFM) has transformed the live imaging of intact biological systems by illuminating only the in-focus plane, dramatically reducing photobleaching and phototoxicity relative to point-scanning approaches \cite{Huisken2004, Keller2008}. Classical selective-plane illumination microscopy (SPIM) architectures have enabled \textit{in toto} imaging of vertebrate development \cite{Keller2010, McDole2018}, large-scale and long-duration cell tracking \cite{Amat2014}, and rapid reconstruction of cellular behaviors in embryogenesis\cite{Krzic2012}. However, these imaging systems typically employ at least two orthogonal objectives, a configuration which forces the use of non-standard sample mounting geometries, resulting in a limited compatibility with many biological contexts of interest, such as \textit{Drosophila} embryos in halocarbon oil, adherent mammalian stem cells in standard culture dishes, or mechanically sensitive three-dimensional organoids. These geometric constraints remain a key bottleneck preventing the application of LSFM to dynamic subcellular phenomena in samples requiring conventional culture conditions.

The distinctive advantages of light-sheet microscopy for live imaging have driven the development of new strategies aimed at improving performance across spatial scales and relaxing sample mounting constraints. Dual-view SPIM (diSPIM) enabled isotropic four-dimensional imaging of embryos and cultured cells \cite{Wu2013}, while axially swept light-sheet microscopy (ASLM) and lattice light-sheet microscopy (LLSM) improved axial resolution and spatiotemporal fidelity in thick tissues \cite{Dean2022,Chen2014}. These light-sheet implementations, however, still rely on at least two objectives positioned in close proximity to the sample, which necessitates specific mounting geometries and often limits their broader adoption. 

Oblique plane microscopy (OPM) provides an elegant solution to the geometric constraints of classical LSFM by generating and imaging a light sheet through a single primary objective in a standard inverted configuration \cite{Dunsby2008}, which finally makes LSFM compatible with any standard sample mounting having a bottom glass coverslip, such as microscope slide, Petri dishes or glass-bottom multiwells plates. The key point is that the excitation laser enters the objective back focal plane off-axis, and this shift results in a tilt of the generated light-sheet on the sample. The emitted light, collected by the same objective, is then relayed into a remote space \cite{Botcherby2007} that compensates optical aberrations before being re-imaged onto the camera. Advances in remote refocusing with optimized relay optics \cite{Yang2019, Alfred2019},  high-speed scanning strategies \cite{Bouchard2015,Voleti2019}, scanned oblique-plane illumination / SOPi-style implementations \cite{Kumar2018,Kumar2019}, and galvanometer-scanned OPM architectures \cite{Sapoznik2020}---have substantially expanded the accessible imaging volume, spatial resolution, and volumetric acquisition rates achievable in single-objective light-sheet systems. More recently, hybrid strategies that combine galvanometer scanning with stage motion have further enlarged the accessible imaging space while preserving single-objective compatibility \cite{Yang2022}. In parallel, scan-lens-free remote-imaging approaches have been introduced to reduce relay complexity and improve photon transfer efficiency in OPM-style geometries \cite{Scandreas, Daetwyler2023}.

 Collectively, these innovations reflect a broad effort to extend single-objective light-sheet microscopy toward higher speed, larger imaging volumes, and improved compatibility with standard specimen geometries. However, despite this progress, many single-objective LSFM implementations remain primarily optimized for qualitative or phenomenological imaging rather than for reproducible quantitative measurements. In particular, quantitative studies of transcriptional activity, chromatin architecture, and single-molecule dynamics are often limited by incomplete or uneven characterization of optical performance, non-deterministic hardware synchronization during high-speed volumetric acquisition, and the lack of an integrated end-to-end analysis pipeline that links raw image data to calibrated biological readouts.

First, the optical geometry of OPM imposes non-trivial constraints on sampling density, PSF anisotropy, and reconstruction accuracy. Although some variation in the effective PSF is intrinsic to tilted-plane imaging, most implementations do not report quantitative characterization across the usable field of view. In our system, we restrict imaging to the diffraction-limited central region and explicitly measure the relevant optical parameters, including transmission efficiency, light-sheet thickness, camera noise characteristics, and flat-field response, as well as chromatic shifts introduced by the remote-refocusing optics. Such characterization is essential for defining valid sampling strategies and for ensuring quantitative interpretation of fluorescence dynamics.

Second, most high-speed volumetric OPM systems rely on software-timed synchronization of galvo scanning, laser excitation, and camera exposure. Software-triggered pipelines accumulate timing jitter and variable latencies, which introduce spatiotemporal distortions and nonuniform exposure. In addition, current OPM implementations typically lack a unified hardware–software control interface, requiring users to coordinate multiple devices independently and further compounding timing variability. These timing errors disproportionately affect quantitative measurements such as transcriptional bursting kinetics, enhancer–promoter correlations, chromatin-locus trajectories, or single-molecule diffusion, where sub-millisecond precision is required.

\begin{figure*}[t!]
  \centering
  \includegraphics[scale=1]{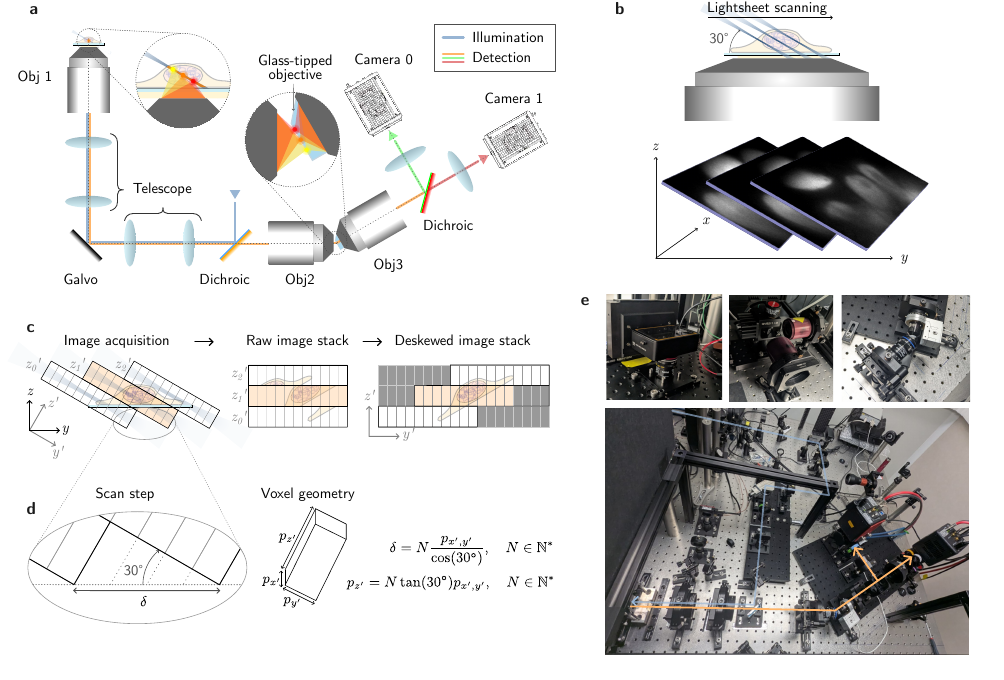} 
  \caption{\textbf{OPM optical layout, geometry, and sampling.}
(\textbf{A}) Simplified schematic of the single-objective oblique plane microscopy (OPM) optical layout. An excitation beam (blue), pre-shaped by cylindrical optics, is delivered through the primary objective (OBJ1, inverted configuration) to generate a planar illumination at an oblique angle $\theta=30^\circ$ with respect to the OBJ1 focal plane. Fluorescence emission (orange) is collected by the same objective and relayed into a remote-refocusing module, where a third objective (OBJ3), tilted by the same angle $\theta$, reimages the oblique plane onto the camera sensor. A dichroic beamsplitter enables simultaneous dual-channel imaging with dedicated cameras.
(\textbf{B}) A galvanometric mirror translates the tilted light sheet laterally through the sample while maintaining a fixed illumination angle. Each camera frame captures a single oblique optical section, and sequential acquisition of adjacent planes enables volumetric imaging.
(\textbf{C}) During volumetric acquisition, fluorescence emission is descanned by the same galvo, resulting in a geometric shear in the raw image stack in which consecutive oblique slices are laterally displaced. Deskewing corrects this shear to reconstruct a geometrically accurate volume, yielding a coordinate system $(x',y',z')$ that is rotated with respect to the laboratory frame $(x,y,z)$.
(\textbf{D}) The discrete sampling of adjacent oblique planes imposes constraints on the galvo step size $\delta$, which must be chosen such that pixels from consecutive frames align after deskewing. This condition relates $\delta$ to the lateral pixel size $p_{x'y'}$ and the illumination angle $\theta$; in our implementation, $p_{x'y'} = 115~\mathrm{nm}$ and $\theta = 30^\circ$, and we select $N = 5$, corresponding to an effective axial sampling of $p_{z'} = 330~\mathrm{nm}$. 
(\textbf{E}) Photographs of the assembled OPM system showing the primary objective with integrated sample incubation chamber (top left), the scanning unit containing the galvo mirror and relay optics (top center), the remote-refocusing module with objectives OBJ2 and OBJ3 (top right), and a top view of the complete microscope assembly (bottom).
}
  \label{fig:setup}
\end{figure*}

Third, although OPM reconstruction methods exist, no implementation provides a comprehensive end-to-end analysis pipeline that converts raw oblique volumes into stable, reproducible biological measurements. Tasks such as nuclei segmentation, single-particle detection, transcription-spot intensity extraction, and 3D trajectory reconstruction are typically handled by fragmented tools with no unified calibration framework. As a result, OPM has not yet reached the level of quantitative robustness needed for mechanistic studies of gene regulation \textit{in vivo}.

Here, we present a fully integrated, quantitatively calibrated oblique plane microscopy platform engineered to address these limitations. Our design combines high numerical aperture remote refocusing with tilt-invariant light-sheet scanning and a hardware-timed control architecture that deterministically synchronizes laser excitation, galvo scanning, and camera exposure, ensuring fixed latencies and reproducible timing across planes, volumes, and long time-lapse acquisitions. We provide a comprehensive optical and detector characterization, including PSF mapping across field and depth, measurement of optical throughput, and a calibrated camera noise model. Together, they establish a photon budget linking illumination parameters to expected signal-to-noise performance.

To support high-throughput volumetric imaging, we developed an integrated acquisition software interface that unifies all device communication within a Python-based framework. This architecture combines user-friendly controls for acquisition parameters, real-time visualization of volumetric data, and high-performance processing that sustains gigabyte-per-second throughput via multiprocessing and shared-memory implementation. For downstream processing, we developed a set of analysis modules for nuclei segmentation, 3D spot detection, sub-pixel Gaussian localization, trajectory inference, and transcriptional activity quantification. Together, these components constitute a unified end-to-end framework that converts raw OPM image stacks into biologically interpretable time series and trajectories.

We demonstrate the capabilities of this platform across three biologically diverse contexts. In \textit{Drosophila} embryos, we image 3D MS2 transcription sites and resolve sister chromatids at active loci, enabling direct measurement of transcriptional bursting dynamics \cite{Bothma2014, Chen2025}. In mouse embryonic stem cells (mESCs), we quantify chromatin-locus dynamics and MS2 bursting, extending classical fixed-cell and 2D live imaging studies to fully three-dimensional volumes \cite{Raj2008, Darzacq2009}. Finally, in mESC-derived gastruloids \cite{vanDenBrink2014, Beccari2018}, we perform 3D transcription imaging in the context of a tissue over extended durations, demonstrating the system's robustness and low phototoxicity in complex multicellular aggregates.

Altogether, this work introduces a fully quantified, end-to-end integrated OPM framework that integrates optical engineering, hardware synchronization, high-throughput computation, and quantitative image analysis. By enabling precise measurements of transcriptional activity and nuclear dynamics across embryos, stem cells, and organoids, this platform provides a versatile foundation for mechanistic studies of gene regulation and tissue self-organization \textit{in vivo}.


\section{Optical layout, geometry, and sampling in OPM}
\label{sec:setup}

Our implementation (Fig.~\ref{fig:setup}A) of oblique plane microscopy (OPM) is based on the original concept introduced by Dunsby~\cite{Dunsby2008} and subsequently extended to achieve  high-NA and  high volumetric speed ~\cite{Botcherby2007, Kumar2018, Yang2019, Voleti2019, Alfred2019, Sapoznik2020}. The core idea of OPM is to generate and image an oblique light sheet through a single primary objective, enabling optical sectioning while preserving full access to the sample on a conventional inverted microscope configuration. This geometry is uniquely suited for imaging \textit{Drosophila} embryos, adherent mammalian stem cells, and three-dimensional organoid systems that cannot be mounted easily in classical LSFM, having dual-objective or more complex configurations.

\blocktitle{Single-objective illumination and fluorescence collection.}
The primary optical element that defines the geometry of the entire microscope is the high–numerical-aperture (NA) objective (OBJ1), which is used for both illumination and fluorescence collection. In our implementation, we employ a silicon oil–immersion objective (NA~1.35), providing high photon-collection efficiency and improved refractive-index matching to biological specimens ($n \approx 1.35$–$1.38$). This index matching minimizes spherical aberrations and preserves image quality at depth, while the working distance remains sufficient to accommodate thick samples such as \textit{Drosophila} embryos and gastruloids.

A high numerical aperture is required not only to achieve high spatial resolution, which is jointly determined by OBJ1 and the remote-refocusing optics described below, but also to permit the lateral offset of the excitation beam relative to the optical axis that is necessary to generate a tilted light sheet without pupil clipping. The excitation beam, pre-shaped by a Powell lens and a series of cylindrical lenses, is therefore delivered through OBJ1 to form a thin light sheet at an oblique angle of $30^\circ$ with respect to the focal plane (Fig.~\ref{fig:setup}B).

The same primary objective collects the emitted fluorescence from the tilted illumination plane. Because this plane is oblique with respect to the camera sensor, direct imaging would place only a narrow line of the plane in focus according to the Scheimpflug principle, while the remainder of the field would be out of focus and increasingly distorted by aberrations for regions above or below the focal plane, as standard microscope objectives obey the sine condition only. To overcome these limitations, the fluorescence is relayed into a remote-refocusing module that optically refocuses the entire oblique plane and reimages it onto the camera sensor at the correct orientation with compensated aberrations, enabling diffraction-limited wide-field detection across the full oblique plane.

\blocktitle{Remote refocusing and relay optics.}
The remote-imaging module follows the aberration-free refocusing principle described by Botcherby \textit{et al.}~\cite{Botcherby2007}. In this configuration, a second objective (OBJ2) forms an intermediate image of the tilted light-sheet plane, and a third objective (OBJ3), tilted with respect to OBJ2 by an angle matching the light-sheet tilt in the sample space, re-images this plane onto the camera sensor (Fig.~\ref{fig:setup}A). The relay lenses between OBJ1 and OBJ2 are chosen to simultaneously satisfy the Abbe sine and Herschel conditions, ensuring diffraction-limited performance across the full extent of the oblique imaging plane. Achieving an aberration-free replica of the fluorescence collected by OBJ1 further requires that the magnification in the remote space, $M_{\mathrm{remote}}$, matches the ratio of the refractive indices of the two focal spaces~\cite{Maxwell1858}; in our implementation, $M_{\mathrm{remote}} = 1.4/1$. The overall system magnification is therefore given by the product $M_{\mathrm{remote}} \times M_{\mathrm{OBJ3}}$.

The optical train incorporates a pair of scan lenses and a galvanometric mirror placed in a plane conjugate to the back focal plane of OBJ1. Driving this galvo with a stair-step waveform changes only the angle at which the excitation beam enters the back focal plane, producing a pure lateral translation of the light sheet through the sample while maintaining a fixed tilt angle~\cite{Kumar2018}. The emitted fluorescence is descanned by the same galvo before entering the remote-refocusing module, such that the intermediate image remains stationary on the camera sensor during volumetric acquisition. Because the illuminated plane intersects the specimen at an angle $\theta$ relative to the OBJ1 focal plane, the sample is spatially sampled along this oblique direction, resulting in an obliquely oriented imaging volume. In the recorded data, galvo descanning leads to a geometric shear in the raw image stack, which must be explicitly corrected during reconstruction. 

\begin{figure*}[t]
  \centering
  \includegraphics[scale=1]{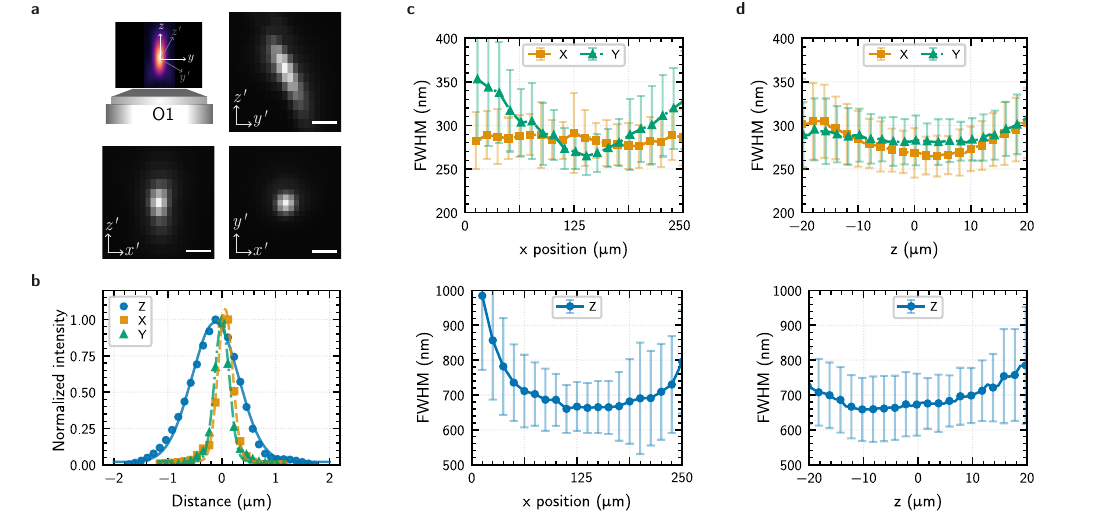}
  \caption{\textbf{PSF and resolution characterization.}
(\textbf{A}) Measured point spread function (PSF) obtained by imaging sub-diffraction fluorescent beads. \textit{Upper left}: the PSF is axially elongated along the laboratory $z$ axis but sampled along the oblique direction $z'$, resulting in a tilted PSF in the rotated coordinate system $(x',y',z')$. \textit{Upper right and bottom}: orthogonal cross-sections $y'z'$, $x'z'$, and $x'y'$ of the measured PSF. Scale bar: $500~\mathrm{nm}$.
(\textbf{B}) Line profiles extracted along the PSF principal axes $(x,y,z)$ with Gaussian fits yield full width at half maximum (FWHM) values of $295 \pm 30~\mathrm{nm}$ along $x$, $305 \pm 30~\mathrm{nm}$ along $y$, and $790 \pm 75~\mathrm{nm}$ along $z$ ($n > 1000$ beads).
(\textbf{C}) Lateral ($x,y$; upper) and axial ($z$; lower) resolution measured as a function of lateral position across the field of view.
(\textbf{D}) Lateral ($x,y$; upper) and axial ($z$; lower) resolution measured as a function of axial position within the sample volume.
Together, these measurements demonstrate diffraction-limited performance over a volumetric field of view of approximately $150 \times 150 \times 40~\mu\mathrm{m}^3$.
}
  \label{fig:psf_resolution}
\end{figure*}

\blocktitle{Tilted-plane geometry and coordinate transformation.}
Because the light sheet is oblique, the imaging coordinate system $(x',y',z')$ defined by the oblique plane does not coincide with the laboratory frame $(x,y,z)$ (Fig.~\ref{fig:setup}C). For an illumination tilt angle $\theta$, the two coordinate systems are related by the following transfomation:
\[
\begin{pmatrix}
x' \\ y' \\ z'
\end{pmatrix}
=
\begin{pmatrix}
1 & 0 & 0 \\
0 & \cos\theta & -\sin\theta \\
0 & \sin\theta & \cos\theta
\end{pmatrix}
\begin{pmatrix}
x \\ y \\ z
\end{pmatrix}.
\]

Galvo scanning and descanning keep the fluorescence image stationary on the camera sensor, such that each acquired frame corresponds to a single oblique $(x',y')$ plane of the sample. While these individual planes are undistorted, stacking them directly along the scan axis produces a volume in which consecutive slices are displaced relative to one another, yielding a skewed representation in the $(y',z')$ plane. To obtain a geometrically correct volume, the data are deskewed by applying an affine shear transformation that re-aligns the oblique slices (Fig.~\ref{fig:setup}C). This operation corresponds to a coordinate transformation only and does not alter the recorded signal, as it does not involve resampling or interpolation. In practice, we perform this deskewing during image analysis rather than during acquisition, as generating deskewed volumes in real time would significantly increase data size. 

The resulting oblique coordinate system defines how camera pixels sample the illuminated plane. We therefore next consider how the choice of magnification and scan parameters sets the effective voxel size in $(x',y',z')$.

\blocktitle{Choice of magnifications and effective voxel size.}
The lateral sampling of the oblique imaging plane is set by the effective magnification in the remote-refocusing space. As described above, this magnification is determined solely by the ratio of refractive indices in the OBJ1 and OBJ2 focal spaces ($n_1/n_2$) and by the magnification of OBJ3. In our configuration, the resulting system magnification of $\sim 56.5\times$ yields an effective pixel size of $p_{x'} = p_{y'} \approx 115~\mathrm{nm}$ in the $(x',y')$ plane, given the camera pixel size of $6.5~\mu$m). This provides appropriate Nyquist sampling relative to the measured in-plane PSF (see Fig.~\ref{fig:psf_resolution}), although the PSF is anisotropic in the laboratory $(x,y,z)$ frame while sampling is performed along $(x',y',z')$. The out-of-plane sampling $p_{z'}$ is set independently by the galvo stair-step amplitude, which determines the spacing between successive oblique slices, and must be chosen to avoid undersampling of the rotated PSF while maintaining hardware-limited volumetric acquisition rates. 

Figure~\ref{fig:setup} summarizes the complete optical path and geometric sampling principles that underlie the quantitative characterization and biological applications described in the sections that follow.


\section{Point-spread function characterization and spatial resolution}
\label{sec:psf}

Quantitative imaging in OPM requires accurate knowledge of the point-spread function (PSF) across the entire tilted imaging plane. Because the illumination and detection paths define an oblique optical section, the PSF geometry differs from that of a conventional widefield or confocal microscope.  In this section, we characterize the lateral and axial resolution of our system using sub-diffraction fluorescent beads, and compare the measured PSF to sampling requirements derived in Sec.~\ref{sec:setup}. 
These measurements follow in the tradition of prior OPM implementations~\cite{Dunsby2008, Botcherby2007, Yang2019, Voleti2019, Sapoznik2020}.

\blocktitle{Experimental measurement of the PSF.}
We imaged 100~nm diameter fluorescent beads embedded in a low density agarose gel, ensuring isolated single-particle images across the entire field of view, and applied the de-skewing method described in Sec.~\ref{sec:setup}. The reconstructed volume was used to extract orthogonal slices through the bead in the $x'y'$, $x'z'$, and $y'z'$ planes (Fig.~\ref{fig:psf_resolution}A). To ensure uniform illumination and detection across wavelengths, PSF measurements were performed independently using 488, 561, and 638~nm excitation, matching the conditions used for the biological experiments presented later.

\blocktitle{Lateral and axial resolution.}
The lateral full-width at half-maximum (FWHM), $\mathrm{FWHM}_{xy}$, was estimated by fitting the transverse bead intensity to a two-dimensional Gaussian model (Fig.~\ref{fig:psf_resolution}B). More complex PSF models (e.g.\ Born–Wolf or vectorial models) were not required because the bead diameter was significantly below the diffraction limit.

\begin{figure*}[t]
  \centering
    \includegraphics[scale=0.8]{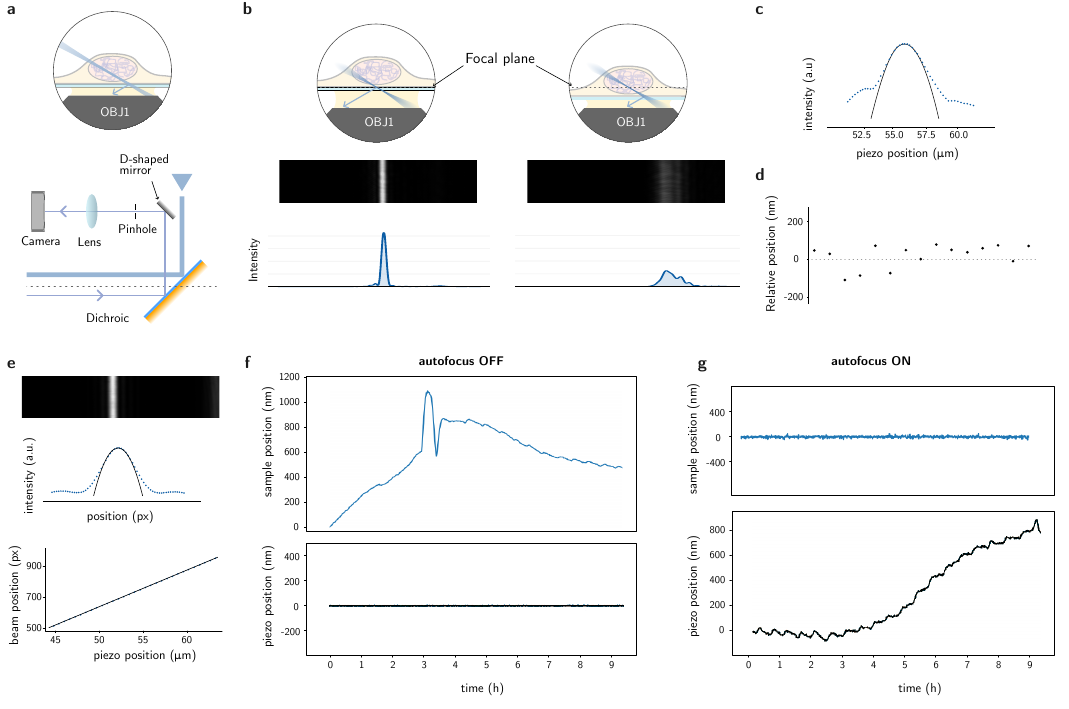}
  \caption{\textbf{System stabilization via back-reflection–based autofocus.}
(\textbf{A}) Schematic of the autofocus optical path. Excitation light reflected at the coverslip–sample interface is redirected by a D-shaped mirror into a dedicated optical arm and imaged onto an auxiliary camera. Axial motion of the sample produces a measurable change in the back-reflection signal.
(\textbf{B}) Determination of the OBJ1 focal plane position from back-reflection intensity. Maximum reflected intensity occurs when the coverslip–sample interface coincides with the focal plane of OBJ1, where the high-NA light sheet reaches its minimal waist. \textit{Left}: interface at the focal plane. \textit{Right}: interface displaced from the focal plane.
(\textbf{C}) Calibration of absolute axial positioning. Integrated back-reflection intensity measured as a function of axial sample position (data points), with a polynomial fit (solid line) used to determine the focal plane location.
(\textbf{D}) Reproducibility of sample repositioning to the focal plane following repeated axial displacements.
(\textbf{E}) Calibration for active drift correction. \textit{Upper}: polynomial fit of the back-reflection signal used to infer relative axial sample position. \textit{Lower}: resulting linear relationship between lateral displacement of the back-reflection spot and axial sample position.
(\textbf{F}, \textbf{G}) Long-term stability during extended acquisitions with autofocus feedback disabled (\textbf{F}) and enabled (\textbf{G}).
}
  \label{fig:stabilization}
\end{figure*}

The axial resolution $\mathrm{FWHM}_{z}$ was obtained by extracting line profiles along the reconstructed $z$-axis and fitting these to a one-dimensional Gaussian. The measured axial resolution is expected to be asymmetric due to the combined effects of oblique illumination and of the overlap of the numerical aperture cones of OBJ2 and OBJ3 in the remote-refocusing unit. These effects have been described qualitatively in prior OPM systems~\cite{Voleti2019, Sapoznik2020}.

Across the central 70\% of the field of view, we obtained lateral resolutions at $488~\mathrm{nm}$ excitation of
\begin{align*}
\mathrm{FWHM}_{x} &= 295 \pm 30~\mathrm{nm}, \\
\mathrm{FWHM}_{y} &= 305 \pm 30~\mathrm{nm}
\end{align*}
and an axial resolution of
\[
\mathrm{FWHM}_{z} = 790\pm 75~\mathrm{nm},
\]
which corresponds to diffraction-limited performance for a single-objective light-sheet detection geometry over a usable field of view of approximately $150 \times 150 \times 40~\mu\mathrm{m}^3$ (Fig.\ref{fig:psf_resolution}C--D).

\section{System stabilization and acquisition reproducibility}
\label{sec:stabilization}

Oblique plane microscopy introduces position-dependent optical effects that arise from its tilted illumination and detection geometry. In particular, axial displacement of the sample relative to the primary objective (OBJ1) focal plane alters the illumination profile of the light sheet and modulates chromatic focal shifts along the detection path. Quantitative imaging therefore requires correction of spatially varying illumination and aberrations, which in turn demands reproducible knowledge of the absolute imaging position within the sample. In live specimens, where fiducial markers are typically unavailable, applying corrections calibrated on external reference samples necessitates a robust strategy for absolute sample positioning and long-term stability.

To address these requirements, we implemented an autofocus and stabilization module that takes advantage of the back-reflection of the excitation beam at the coverslip–specimen interface. This approach provides absolute axial positioning relative to the OBJ1 focal plane and enables compensation of mechanical instabilities (e.g. thermal drift or vibrations) during extended acquisitions. Because the method operates independently of the detection path, it can be readily integrated into existing OPM implementations without modifying fluorescence imaging optics.

\blocktitle{Implementation.}
As described in Sec.~\ref{sec:setup}, the excitation beam enters OBJ1 with a lateral offset to generate the oblique illumination plane. A small fraction of the incident light undergoes Fresnel reflection at the coverslip–specimen interface and exits OBJ1 through the opposite side of the back aperture. A D-shaped mirror positioned upstream of the main dichroic redirects this back-propagating light into a dedicated stabilization arm where a $1~\mathrm{mm}$ pinhole isolates the coverslip reflection from spurious optical reflections (Fig.\ref{fig:stabilization}A). Additional relay optics project the isolated signal onto a camera positioned at a sample-conjugate plane. Axial displacement of the sample relative to OBJ1 produces a corresponding lateral displacement of the back-reflection spot on the stabilization camera chip.

\begin{figure*}[h]
  \centering
  \includegraphics[scale=0.8]{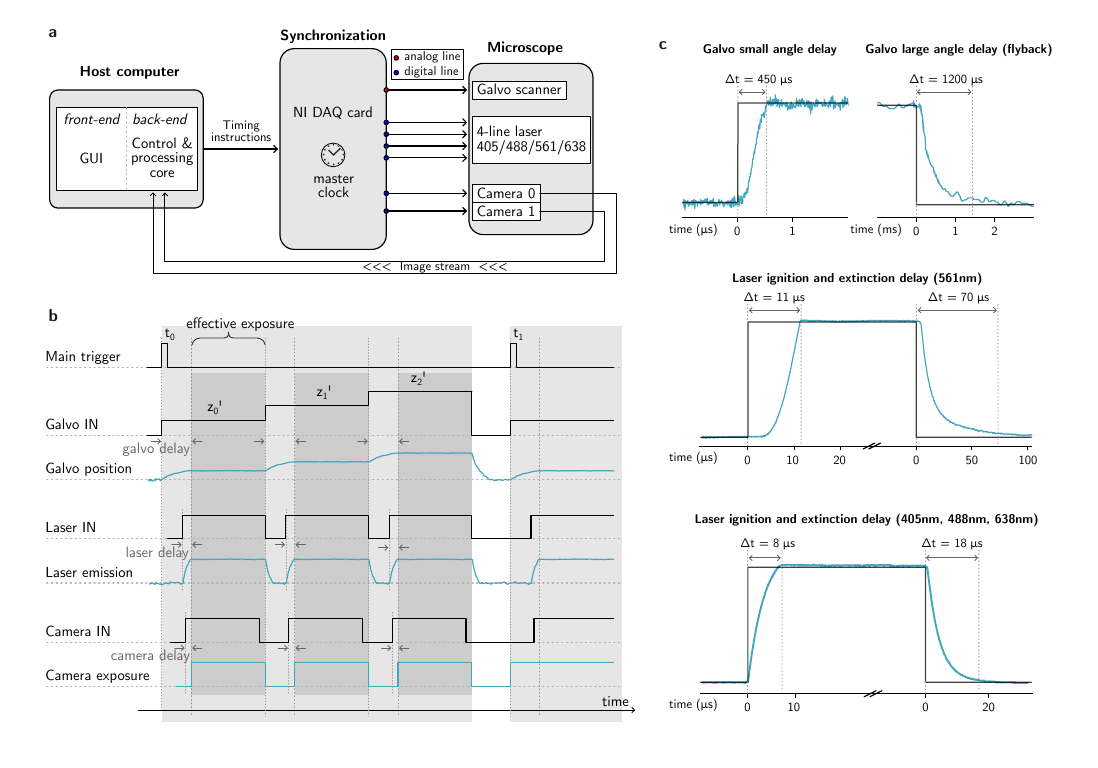}
  \caption{\textbf{Hardware synchronization of galvo, lasers, and camera.}
  (\textbf{A}) Hardware architecture. A DAQ card functions as the master timing controller, generating clock-synchronized analog (galvo) and digital (camera, laser) signals. Precomputed timing sequences are loaded onto the DAQ for autonomous, deterministic execution during acquisition, ensuring fixed relative timing between galvo motion, laser illumination, and camera exposure. The host computer asynchronously acquires frames for visualization and processing.
  (\textbf{B}) Timing diagram for one volume acquisition. Command signals (black) sent to galvo (analog staircase), camera (exposure gates), and laser (illumination pulses) are shown with measured device responses (blue). Finite latencies between commands and responses require compensation to synchronize illumination with camera exposure during galvo dwell periods.
  (\textbf{C}) Measured device latencies. Quantified response times for galvo (mechanical settling), camera (trigger delay), and laser (switching time) enable compensation in the DAQ timing sequences, ensuring optimal synchronization and acquisition efficiency.}
  \label{fig:synchronization}
\end{figure*}

\blocktitle{Sample repositioning.}
The sample repositioning routine establishes a defined axial reference relative to the OBJ1 focal plane, providing a reproducible coordinate system for position-dependent corrections. During sample preparation, coverslip mounting typically introduces a small tilt that varies between specimens, rendering the lateral position of the back-reflection spot insufficient for absolute positioning. Instead, we use the integrated back-reflection intensity, which reaches a maximum when the coverslip surface coincides with the OBJ1 focal plane, where the light-sheet waist is minimal (Fig.\ref{fig:stabilization}B). This maximum is identified by axially scanning the piezo stage and fitting the resulting intensity profile with a second-order polynomial (Fig.\ref{fig:stabilization}C). The sample is then displaced by a predefined axial offset to center the imaging region symmetrically about the light-sheet waist. Using a light-sheet numerical aperture of $\mathrm{NA}=0.3$, this procedure yields repositioning precision of $\pm 80~\mathrm{nm}$ (Fig.\ref{fig:stabilization}D) and is robust to sample-to-sample variations.

\blocktitle{Compensation of mechanical instabilities.}
While the back-reflection intensity provides absolute axial referencing, the lateral position of the back-reflection spot reports relative axial displacements during acquisition. A calibration procedure (Fig.~\ref{fig:stabilization}E) establishes a linear relationship between lateral spot displacement and axial sample motion. During time-lapse imaging, the system periodically measures the back-reflection position and compares it to the reference value established at the start of the experiment. Deviations are corrected by adjusting the sample stage using the calibrated conversion factor, restoring the imaging plane to its target position. Each correction cycle requires approximately $500~\mathrm{ms}$ and maintains sub-micrometer axial stability; over a 12-hour acquisition, sample position drift remained within $\pm 25~\mathrm{nm}$ (Fig.\ref{fig:stabilization}F--G), effectively suppressing fluctuations of the sample/OBJ1 positioning due to thermal variations or other sources of mechanical instability.

Together, these stabilization procedures ensure that position-dependent optical corrections remain valid throughout long-term experiments and across independent acquisitions. By enforcing reproducible axial positioning and suppressing mechanical drifts, the autofocus module provides a stable spatial reference frame that complements the optical characterization described above and supports reliable, quantitative time-resolved imaging.


\section{Hardware-timed synchronization}
\label{sec:sync}

Timing precision is a fundamental requirement for high-rate volumetric imaging, as it directly impacts both acquisition efficiency and quantitative accuracy. In OPM, each volumetric acquisition consists of a sequence of oblique planes sampled at defined galvanometric mirror positions, with laser illumination and camera exposure accurately synchronized. Over long time-lapse experiments, even small timing errors or cumulative latency can distort the effective time axis, introduce spatial misregistration between planes, or lead to frame loss. Accurate quantitative measurements therefore require both a detailed characterization of device-specific latencies and a deterministic synchronization strategy that remains stable over thousands of volumes.

\begin{figure*}[t]
  \centering
  \includegraphics[scale=0.8]{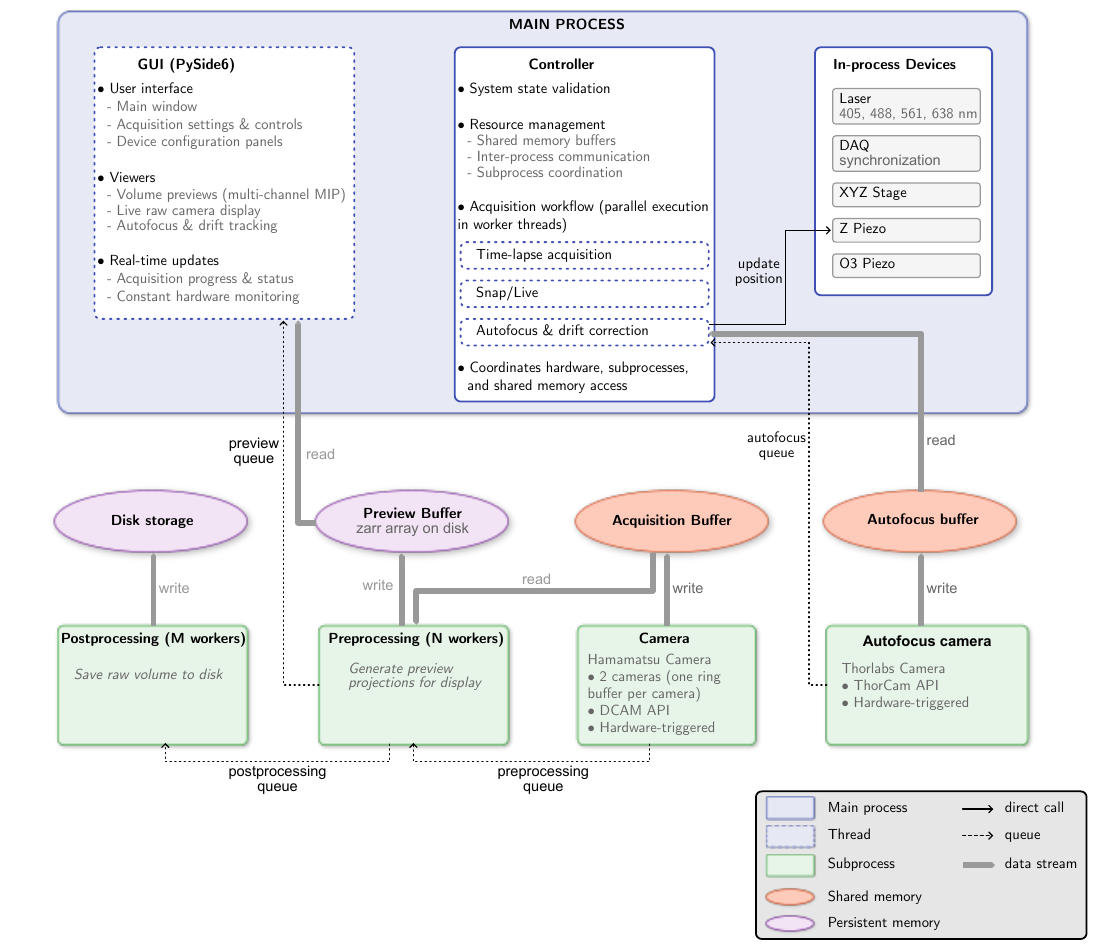}
  \caption{\textbf{Software pipeline for high-throughput volumetric imaging.}
Schematic of the process-based software architecture used for acquisition, visualization, and data handling. Hardware control is managed by a main process, while camera readout, preprocessing, and postprocessing are executed in independent subprocesses. Data flow proceeds as follows: (\textbf{1}) camera subprocesses acquire image frames into shared circular buffers and post volume-completion notifications to message queues; (\textbf{2}) preprocessing worker pools monitor these queues, retrieve buffer metadata, and generate maximum-intensity projections that are written to Zarr arrays; (\textbf{3}) the graphical user interface reads the Zarr arrays to enable real-time visualization during acquisition; (\textbf{4}) postprocessing workers receive queued tasks and write raw volumetric data to disk. Queue-based messaging ensures process independence, while shared-memory buffers enable zero-copy data access, supporting sustained high-throughput acquisition without software bottlenecks.
}
  \label{fig:software}
\end{figure*}

\blocktitle{Deterministic hardware synchronization.}
We implemented a fully hardware-triggered acquisition scheme based on a data-acquisition (DAQ) card that serves as the central timing controller. As summarized in Fig.~\ref{fig:synchronization}A–B, timing sequences corresponding to one complete volume are precomputed on the host computer and loaded onto the DAQ prior to imaging. During acquisition, the DAQ autonomously executes these sequences, triggering successive volumes at precisely defined intervals and ensuring deterministic timing over long time-lapse experiments spanning thousands of volumes.

Within each volume, the DAQ card coordinates galvanometric mirror positioning, laser emission, and camera integration using synchronized analog and digital outputs derived from a single phase-locked master clock. The host computer operates asynchronously, capturing camera frames for real-time visualization and downstream processing (Sec.~\ref{sec:software}). Although this approach reduces the flexibility inherent to software-timed acquisition, it provides the deterministic timing necessary for accurate time-resolved measurements in high-speed volumetric imaging.

\blocktitle{Device latencies and response characterization.}
Camera exposure and laser emission are controlled via TTL signals, with the HIGH state defining the integration and illumination intervals. The galvanometric mirror is driven by an analog stair-step waveform, in which each voltage level corresponds to a discrete angular position defining a specific oblique imaging plane. All devices exhibit finite and reproducible response delays due to signal propagation and electronic switching, typically on the order of tens of microseconds. The galvo introduces additional latency associated with mechanical inertia, which can reach several hundred microseconds depending on the angular step size.

To compensate for these effects, we directly measured device-specific response times (blue traces in Fig.~\ref{fig:synchronization}B–C) and incorporated the corresponding delays into the command timing sequences (black traces). This ensures that laser illumination occurs exclusively during the camera integration window while the galvo remains stationary at the target position for each plane. As a result, photon collection efficiency is maximized across the entire volume, dead time between frames is minimized, and the achievable volumetric acquisition rate is not limited by synchronization artifacts.


\section{Software acquisition pipeline and reconstruction throughput}
\label{sec:software}

OPM imaging generates data streams exceeding $1~\mathrm{GB/s}$ per camera, which can rapidly bottleneck acquisition if data are not processed and offloaded in real time. Unmanaged data accumulation leads to memory exhaustion and acquisition failure, particularly during long time-lapse experiments. To address these demands, we developed a custom Python-based software framework that controls each device and modality (real-time acquisition, processing, visualization, and storage of volumetric data) without requiring to alternate between the proprietary software of the devices producers or tertiary software to perform specific functions.

Our implementation builds on the SOLS Python control framework \cite{Alfred2019}, which we extensively adapted to accommodate the specific hardware configuration and timing requirements of our OPM implementation. For the user interface, we adopted the graphical layout of the Brighteyes-MCS platform as a design template and customized it to the specific needs of our system. The backend control, acquisition, and synchronization logic were developed independently and do not rely on existing Brighteyes-MCS functionality. This separation between backend control and frontend layout provides both low-level flexibility and ease of use, enabling routine operation by users without extensive programming experience.

\blocktitle{Python software architecture.}
The software adopts a modular, process-based architecture that decouples hardware control from computationally intensive data processing. Each hardware component is managed by a dedicated device module that encapsulates device-specific communication protocols. Most device modules operate within the main process; camera modules, however, run in dedicated subprocesses due to the substantial computational load associated with continuous frame acquisition and buffering. This design circumvents Python’s Global Interpreter Lock and enables true parallel execution at the cost of increased inter-process communication complexity.

At startup, the main process launches the GUI and instantiates all device modules, establishing hardware communication and configuration interfaces. Once acquisition parameters are defined, timing sequences are computed and transferred to the DAQ, and a validation routine verifies device readiness and configuration integrity. During acquisition, preprocessing and postprocessing worker pools are spawned, and shared circular buffers are allocated to enable parallel data handling. Software-level device communication is suspended throughout acquisition, with all timing governed exclusively by DAQ-generated hardware triggers.

\blocktitle{Real-time data processing.}
The acquisition pipeline follows a shared-memory producer--consumer design. Camera subprocesses continuously poll for incoming frames and transfer data from camera buffers into shared acquisition buffers. Upon completion of a volumetric stack, preprocessing workers are notified via inter-process queues and generate maximum-intensity projections along the principal axes. These projections are written to disk-backed Zarr arrays and made available to the GUI for real-time visualization. Subsequently, a postprocessing worker pool asynchronously saves the raw volumetric data to disk. This pipeline operates at hardware-limited speeds, preventing software-induced bottlenecks during sustained high-throughput acquisition.

\blocktitle{Graphical user interface.}
An integrated GUI provides streamlined access to all hardware and acquisition parameters, facilitating routine operation of the microscope. 
The interface supports the full imaging workflow, including autofocus routines, acquisition-parameter optimization, and pre-acquisition visualization. During imaging, the GUI continuously displays maximum-intensity projections retrieved from the shared Zarr arrays, providing real-time feedback on acquisition progress and data quality.

Together, this software architecture ensures that data acquisition, processing, and storage keep pace with hardware-limited volumetric imaging rates without introducing computational bottlenecks. By decoupling real-time acquisition from downstream processing and visualization, the pipeline preserves the deterministic timing enforced by hardware synchronization (Sec.~\ref{sec:sync}) while enabling immediate feedback and scalable data handling.


\section{Quantitative transcription imaging in \textit{Drosophila} embryos}
\label{sec:fly}

\begin{figure*}[t]
  \centering  
  \includegraphics[scale=1]{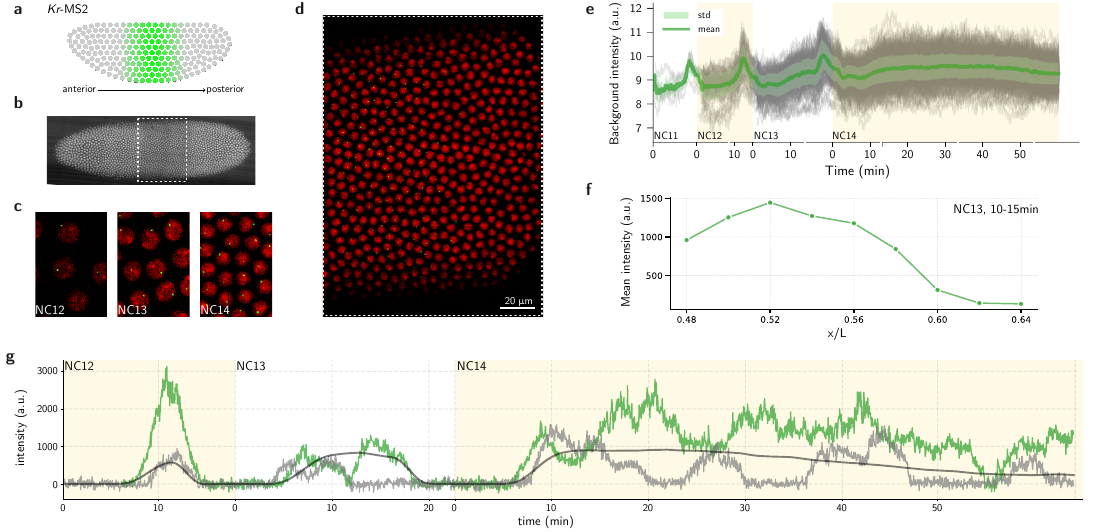}
  \caption{\textbf{Quantitative transcription imaging in fly embryos.}
(\textbf{A}) Schematic representation of the \textit{Krüppel} (\textit{Kr}) expression pattern along the anterior–posterior axis of a blastoderm stage \textit{Drosophila} embryo.
(\textbf{B}) Wide-field embryo image with fluorescently marked surface nuclei (Histone-RFP label) at the end of a long-term time-lapse experiment (3h-post-oviposition), with the dashed rectangle indicating the imaged field of view.
(\textbf{C}) Time-lapse projections of a subregion of the imaged field of view at three successive nuclear cycles: 12, 13, and 14. Nuclei are fluorescently labeled in red, while active transcription sites of the \textit{Kr} gene are visualized using the MS2/MCP–GFP system (see Methods \textcolor{red}{[???]}).
(\textbf{D}) Projection of the full imaged field of view at the end of the time-lapse experiment. 
(\textbf{E}) Background MCP–GFP fluorescence intensity measured over an extended time-lapse spanning multiple nuclear cycles, demonstrating low photobleaching and phototoxicity during high-speed volumetric imaging.
(\textbf{F}) Measured \textit{Kr} expression profile along the anterior–posterior axis during nuclear cycle 13, illustrating the quantitative fidelity of the acquired transcriptional readouts.
(\textbf{G}) Representative transcriptional intensity traces from individual nuclei, showing temporal dynamics of \textit{Kr} transcription across multiple nuclear cycles.
Panels in this figure focus on MS2-labeled transcription sites and quantitative transcriptional intensity measurements.
}
  \label{fig:fly_app}
\end{figure*}

To demonstrate the capabilities of the system for quantitative live imaging, we applied OPM to transcriptional dynamics in developing \textit{Drosophila} embryos. Early embryogenesis provides an ideal test case: rapid nuclear divisions (cycle duration $\sim$10--20~min), stereotyped spatial patterning of gene expression, and the availability of MS2/MCP fluorescent tagging for single-locus visualization combine to impose stringent demands on imaging speed, photon efficiency, and long-term stability. Moreover, the established quantitative framework for developmental transcription in this system enables direct validation of our measurements against prior benchmarks.

We focused on the gap gene \textit{Krüppel} (\textit{Kr}), which exhibits a well-characterized spatial expression domain along the anterior--posterior axis (Fig.~\ref{fig:fly_app}A) and a consistent temporal activation during nuclear cycles 12--14. Embryos were imaged in a $100 \times 120 \times 20~\mu\mathrm{m}^3$ volume at a volumetric rate of 0.5~Hz over two hours, spanning multiple nuclear cycles (Fig.~\ref{fig:fly_app}B--D).

\blocktitle{Long-term imaging with minimal photodamage.}
High-speed volumetric imaging requires sustained illumination over thousands of acquisition cycles, raising concerns about photobleaching and phototoxicity. To assess imaging-induced perturbations, we quantified background MCP--GFP fluorescence throughout extended time-lapse acquisitions. Over three hours of continuous imaging at 0.5~Hz ($>4000$ volumes), background fluorescence remained stable (Fig.~\ref{fig:fly_app}E), and embryos progressed through normal nuclear divisions. These observations indicate that the imaging conditions enable long-term volumetric acquisition without detectable photobleaching or developmental arrest.

\blocktitle{Quantitative analysis of transcriptional dynamics.}
The full imaged volume (Fig.~\ref{fig:fly_app}D) captures hundreds of nuclei simultaneously, enabling both population-level and single-nucleus analysis. Spatial validation confirms measurement fidelity: the extracted MS2 fluorescence profile during nuclear cycle 13 (Fig.~\ref{fig:fly_app}F) exhibits the characteristic \textit{Kr} expression domain centered at approximately 45--55\% embryo length, consistent with previous measurements~\cite{Chen2025}. Temporal analysis at the single-nucleus level (Fig.~\ref{fig:fly_app}G) reveals stochastic transcriptional dynamics, with intensity traces exhibiting bursts, plateau phases, and mitotic terminations that can be interpreted using promoter-state models. The ability to extract hundreds of such traces from defined spatial positions within a single embryo enables statistical inference of transcriptional parameters and their spatial variation along the morphogen gradient. Together with complementary single-molecule mRNA tracking (Supplementary Sec.~S4), these measurements illustrate the suitability of OPM for quantitative studies of transcriptional regulation in \textit{Drosophila} embryos.




\section{Quantitative chromatin and transcription dynamics in cultured mammalian embryonic stem cells}
\label{sec:mesc}

\begin{figure*}[t]
  \centering
  \includegraphics[scale=1]{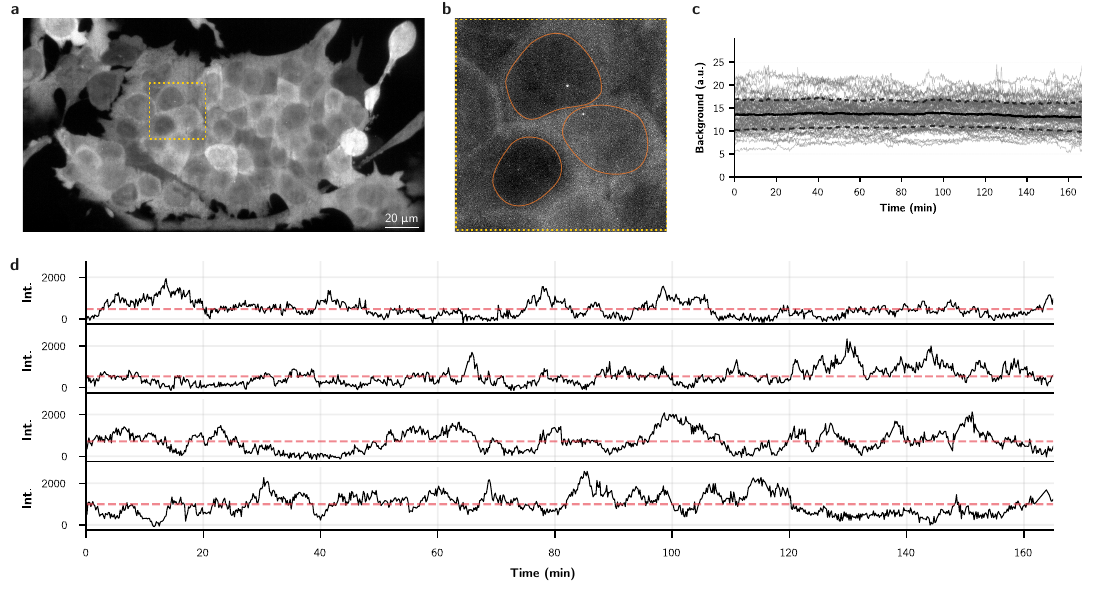}
  \caption{\textbf{Quantitative transcription imaging in mouse embryonic stem cells.} 
(\textbf{A}) Projection of a mouse embryonic stem cell colony imaged using OPM.
(\textbf{B}) Magnified view of the dashed region in panel A, showing segmented nuclei (orange outlines) and active \textit{Sox2} transcription sites labeled using the MS2/MCP–GFP system.
(\textbf{C}) Time evolution of the MCP–GFP background fluorescence intensity during extended volumetric imaging, indicating low photobleaching and phototoxicity for moderately fluorescent specimens.
(\textbf{D}) Representative transcriptional intensity traces from individual cells, illustrating burst-like \textit{Sox2} transcription dynamics with alternating active and inactive periods. The red dashed line indicates the mean transcriptional activity over the imaging session.
Representative panels focus on MS2-labeled transcription sites and nuclear-scale transcription dynamics; extended examples are provided in Supplementary Sec.~S4.
}
  \label{fig:mesc_app}
\end{figure*}

We next applied OPM to mouse embryonic stem cells (mESCs), which present distinct imaging challenges compared to \textit{Drosophila} embryos, including weaker transcriptional signals and slower temporal dynamics. mESCs provide a powerful model for studying chromatin organization, transcription, and nuclear architecture in a pluripotent mammalian context, yet quantitative live imaging in these systems has traditionally been limited by phototoxicity during extended three-dimensional acquisitions. A key advantage of the single-objective OPM geometry is its compatibility with standard cell-culture formats: coverslip-bottom dishes can be imaged directly on the primary objective within an environmental chamber, enabling observation of cells under conventional \textit{in vitro} conditions with regulated temperature, humidity, and CO\textsubscript{2} concentration.

\blocktitle{Photostability and phototoxicity in long-term imaging.}
We imaged mESC colonies expressing MS2-labeled \textit{Sox2} transcripts and nuclear markers in volumes of $150 \times 100 \times 20~\mu\mathrm{m}^3$ at a volumetric rate of one volume every 10~s for more than two hours (Fig.~\ref{fig:mesc_app}A--B). To assess photostability under these conditions, we monitored background MCP--GFP fluorescence intensity throughout the acquisitions (Fig.~\ref{fig:mesc_app}C). The cytoplasmic MCP--GFP pool remained stable over extended imaging sessions, with minimal photobleaching despite continuous volumetric illumination. These observations indicate that light-sheet excitation enables long-term imaging of moderately fluorescent mammalian specimens without detectable loss of fluorophore signal or adverse effects on cell viability.

\blocktitle{Single-cell transcriptional activity.}
Transcriptional intensity traces extracted from individual cells (Fig.~\ref{fig:mesc_app}D) demonstrate that OPM resolves \textit{Sox2} expression dynamics at single-locus sensitivity in adherent mESCs. The traces reveal substantial cell-to-cell variability in both baseline activity, indicated by the red dashed mean line, and temporal dynamics, reflecting the intrinsic transcriptional heterogeneity of stem cell populations. Despite weaker fluorescence signals and a lower signal-to-noise ratio than in embryonic systems, individual transcriptional events remain clearly distinguishable from background, validating the suitability of the platform for quantitative transcription measurements under standard mammalian cell culture conditions.


\section{Quantitative imaging in mammalian 3D culture systems}
\label{sec:gastruloids}

\begin{figure*}[t]
  \centering
  \includegraphics[scale=0.9]{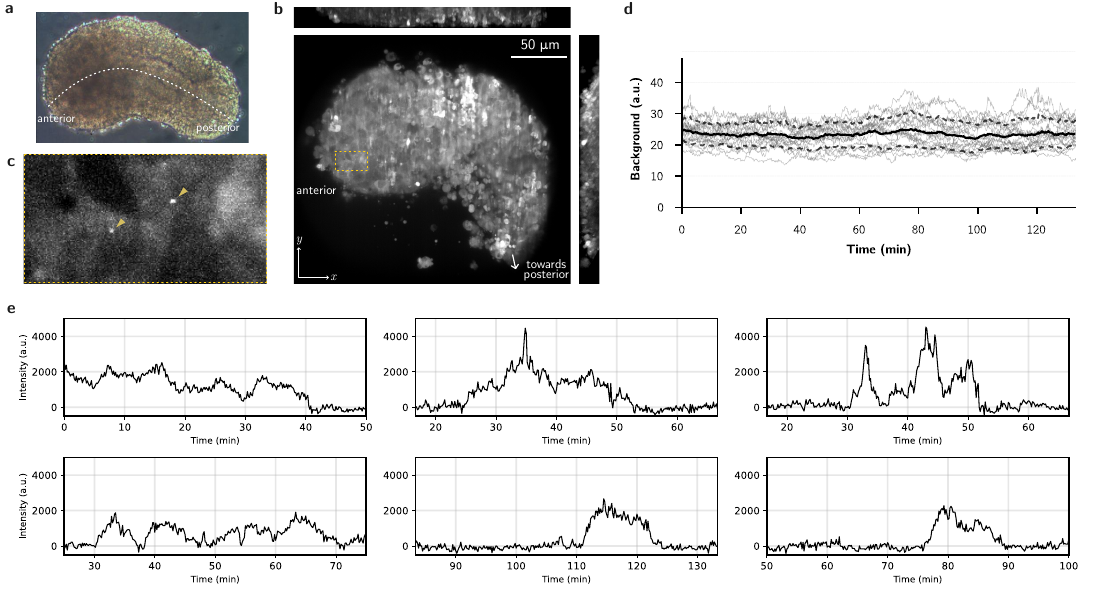}
  \caption{\textbf{Quantitative transcription imaging in gastruloids.}
(\textbf{A}) Wide-field image of a whole gastruloid at 120~h of the differentiation protocol, with the anterior–posterior axis indicated.
(\textbf{B}) Volumetric projections of a gastruloid imaged using OPM, showing the GFP signal composed of diffuse MCP–GFP background fluorescence with superimposed \textit{Sox2} transcription sites.
(\textbf{C}) Magnified view of the dashed region in panel B, highlighting individual \textit{Sox2} transcription spots.
(\textbf{D}) Time evolution of the MCP–GFP background fluorescence measured at different positions within the gastruloid, indicating low photobleaching and phototoxicity during extended volumetric imaging in a thick, optically dense specimen.
(\textbf{E}) Representative transcriptional intensity traces from individual \textit{Sox2} transcription sites, showing that burst-like dynamics are maintained within the three-dimensional gastruloid context.
Panels focus on MS2-labeled transcription imaging and quantitative transcriptional intensity measurements in three-dimensional multicellular systems. 
}
  \label{fig:mesc_gastruloid_app}
\end{figure*}

We next applied OPM to mESC-derived gastruloids, which recapitulate key aspects of early embryonic development and represent the most demanding imaging conditions among the systems examined here. Compared to monolayer cell cultures, multicellular aggregates are optically heterogeneous and exhibit substantial light scattering, which degrades image quality with depth and poses challenges for quantitative live imaging. Gastruloids were imaged at 120~hours post-aggregation, when anterior–posterior axis specification is established (Fig.~\ref{fig:mesc_gastruloid_app}A). Focusing on the anterior pole, where \textit{Sox2} transcription is active (Fig.~\ref{fig:mesc_gastruloid_app}B--C), we acquired volumetric datasets spanning $200 \times 200 \times 30~\mu\mathrm{m}^3$ every 10~s for three hours.

\blocktitle{Photostability and long-term three-dimensional imaging.}
Thick, scattering specimens exhibit spatially varying optical properties that can lead to heterogeneous illumination and position-dependent photodamage. To assess imaging-induced perturbations under these conditions, we monitored background MCP--GFP fluorescence at multiple positions within gastruloids during extended acquisitions (Fig.~\ref{fig:mesc_gastruloid_app}D). Fluorescence levels remained stable across all sampled locations despite optical heterogeneity, indicating low photobleaching and phototoxicity throughout the imaged volume.

\blocktitle{Single-cell transcriptional dynamics in a multicellular context.}
Representative transcriptional intensity traces from individual \textit{Sox2} transcription sites (Fig.~\ref{fig:mesc_gastruloid_app}E) demonstrate that burst-like transcription dynamics are preserved within the three-dimensional gastruloid environment. These measurements show that OPM resolves transcriptional activity at single-cell resolution in dense, self-organizing multicellular structures. The ability to quantify single-locus transcriptional dynamics in such optically complex systems highlights the suitability of OPM for quantitative studies of gene expression during mammalian developmental self-organization.



\section*{Discussion}
\label{sec:discussion}

The single-objective OPM implementation presented here provides a quantitatively characterized and experimentally robust platform for high-speed volumetric imaging across a wide range of biological systems. By combining high NA optics, tilt-invariant remote refocusing, well-calibrated sampling geometry (Sec.~\ref{sec:setup}), PSF characterization (Sec.~\ref{sec:psf}), precise hardware-timed synchronization (Sec.~\ref{sec:sync}), and high-throughput acquisition software (Sec.~\ref{sec:software}), the system delivers consistent, reproducible performance suitable for demanding quantitative applications. These capabilities distinguish OPM from many existing light-sheet implementations \cite{Kim2023}, positioning it as a powerful tool for studying transcription and nuclear organization \textit{in vivo}.

A central conceptual contribution of this work is the transformation of OPM from a primarily descriptive imaging modality into a fully quantitative measurement instrument \cite{Glaser2016}. This system is calibrated at every level of the imaging chain: sampling geometry, PSF shape and field dependence, synchronization, and computational throughput. Such end-to-end calibration is rare among single-objective light-sheet implementations, providing defined spatial and temporal performance metrics that can be reproduced across sessions, samples, and experimental conditions \cite{Stelzer2021}.

Two important aspects of quantitative imaging deserve particular emphasis: flat-field correction and chromatic-aberration correction. Both are essential for precise intensity-based measurements, but both are substantially more challenging in OPM than in conventional geometries. Flat-field correction must account not only for illumination non-uniformity \cite{Baumgart2012}, but also for the projection and shear introduced by the oblique plane, which cause position-dependent intensity distortions not present in standard widefield or confocal imaging. Chromatic aberration correction is likewise nontrivial: the different illumination and detection paths for each wavelength, combined with the tilt of the imaging plane, produce wavelength-dependent shifts that vary across the field of view. Accurate multi-color registration therefore requires wavelength-specific affine transforms and calibration procedures similar to those developed for multichannel single-molecule imaging. While not yet fully generalized here, these corrections are critical foundations for future quantitative extensions.

Looking ahead, several technical extensions could further expand the capabilities of OPM. Two-photon excitation represents a promising means of increasing imaging depth in scattering tissues while maintaining single-objective accessibility, and early demonstrations of two-photon OPM implementations have been reported \cite{Bouchard2015, Galwaduge2016, Voleti2019}. Adaptive optics provides a natural upgrade for correcting sample-induced aberrations in organoids and embryonic tissues \cite{Ji2017}. Multi-view or hybrid illumination strategies could enhance isotropy and reduce residual blur in deep structures while remaining compatible with a single-objective configuration. Each of these directions is feasible within the existing framework but will require careful stabilization and quantitative calibration comparable to that performed here.

In summary, this work establishes oblique plane microscopy as a robust and quantitatively calibrated single-objective light-sheet platform for dynamic transcription imaging in complex living systems. By integrating high–numerical-aperture optical design, tilt-invariant light-sheet scanning, deterministic hardware-timed synchronization, and high-throughput acquisition software within a single framework, we transform OPM from a primarily descriptive imaging modality into a quantitative measurement instrument. The resulting platform enables reproducible three-dimensional imaging of MS2-labeled transcription sites in intact \textit{Drosophila} embryos, cultured mouse embryonic stem cells, and self-organizing gastruloids, linking optical performance directly to biologically interpretable transcriptional readouts. Beyond these demonstrations, the modular and fully characterized nature of the system provides a foundation for future extensions, such as multi-color transcription imaging, chromatin locus tracking, and deeper imaging in scattering tissues, while maintaining compatibility with standard sample geometries. As such, this work positions single-objective OPM as a versatile and quantitative tool for investigating gene regulation and nuclear dynamics across developmental and stem-cell models.


\section*{Methods}

\blocktitle{Microscope description}
The microscope is based on a single-objective oblique plane microscopy (OPM) architecture, adapted and extended from the implementation described in Ref.~\cite{Sapoznik2020}. A high–numerical-aperture silicone oil–immersion objective (Olympus UPLSANSAPO 100$\times$, NA~1.35, working distance 0.2~mm) serves as the primary objective (OBJ1) for both illumination and fluorescence collection in an inverted configuration.

Fluorescence collected by OBJ1 and tube lens TL1 (Olympus SWTLU-C) is relayed through a pair of achromatic scan lenses (Thorlabs CLS-SL), with a galvanometric mirror (Thorlabs GVS011) positioned in a plane conjugate to the back focal plane of OBJ1. A second objective (OBJ2; Olympus UPLXAPO 40$\times$, NA~0.95, working distance 0.18~mm), combined with a tube lens (TL2, effective focal length 321~mm~\cite{Alfred2019}), forms an intermediate image in the remote space. The relay optics are designed to simultaneously satisfy the Abbe sine and Herschel conditions, ensuring aberration-free remote refocusing. A third objective (OBJ3; Special Optics AMS-AGY v1.1), tilted at the same angle of the light-sheet illumination on the sample, forms with a tube lens TL3 (Nikon MXA20696) an undeformed image on the camera chip. OBJ3 is mounted on a piezoelectric actuator (Edmund Optics 85-008, travel range 100~$\mu$m), which enables fine axial alignment of the remote imaging plane.

To increase imaging throughput, a long-pass dichroic beamsplitter (Semrock Di03-R561-t3-25$\times$36), placed between OBJ3 and TL3, separates the emitted fluorescence onto two channels that are imaged by two sCMOS cameras (Hamamatsu C14440-20UP and C15440-20UP), which operate in parallel. Longer-wavelength emission is transmitted to one camera and shorter-wavelength emission is reflected to the other. Camera-specific emission filters are placed immediately upstream of each detector, and both dichroics and emission filters are interchangeable to accommodate different fluorophore combinations.

Excitation light is provided by a four-line solid-state laser (Cobolt Skyra, 405/488/561/638~nm) delivering pre-aligned and collimated beams. The excitation beam is spatially filtered using a single-mode fiber (Thorlabs S405-XP) and collimated with an achromatic fiber-port collimator (Thorlabs PAF2-A4A). The beam is expanded using two cylindrical-lens telescopes that independently control the vertical and horizontal beam dimensions. The first telescope, consisting of a Powell lens (LaserLine Optics LOCP-8.9R30-1.0) and an achromatic cylindrical lens (Thorlabs ACY254-050-A, focal length 50~mm), produces vertical elongation with a homogeneous intensity profile. The second telescope, composed of achromatic cylindrical lenses (Thorlabs ACY254-100-A and ACY254-250-A; focal lengths 100~mm and 250~mm), provides horizontal expansion with a magnification of $2.5\times$.

The expanded beam is mounted on a translation stage (Thorlabs PT1/M with MB3090/M bracket) to enable controlled lateral displacement, which sets the oblique illumination angle in the sample space. The beam is focused by a cylindrical lens (Thorlabs ACY254-100-A, focal length 100~mm) onto a resonant scanner (Cambridge Technology CRS, 12~kHz), positioned in a plane conjugate to the front focal plane of OBJ1, enabling pivoting illumination to minimize stripe artefacts~\cite{Huisken2007}. The beam is subsequently relayed by a spherical lens (Thorlabs AC254-100-A, focal length 100~mm) into the main detection path. An adjustable slit (Thorlabs VA100), positioned in a plane conjugate to the back focal plane of OBJ1, controls the effective light-sheet numerical aperture $\mathrm{NA}_{\mathrm{LS}}$, and thus the light-sheet thickness and Rayleigh range.

A dichroic beamsplitter (Semrock Di03-R405/488/561/635-t3-25$\times$36), placed between TL2 and OBJ2, couples the excitation and detection paths by reflecting the excitation beam toward OBJ1 while transmitting the collected fluorescence toward the remote-refocusing optics. The shaped beam is delivered to OBJ1 with a controlled lateral offset relative to the optical axis, generating a tilted illumination plane at an angle $\theta = 30^\circ$ with respect to the OBJ1 focal plane.

\blocktitle{Imaging of \textit{Drosophila} embryos}
For two-color fluorescence imaging, His-RFP, MCP-GFP virgin females were crossed with males carrying CRISPR-engineered MS2 loops inserted into the gene of interest, here \textit{Krüppel} (\textit{Kr}). The resulting embryos exhibited fluorescently labeled nuclei (His-RFP) and MS2-tagged transcription sites (MCP-GFP). Embryos were manually dechorionated and mounted between a 1.5H glass coverslip and an oxygen-permeable film (Lumox 25, Sarstedt), then immersed in Halocarbon~27 oil. The oil provides the oxygen permeability required for embryonic development and a refractive index ($n \approx 1.41$) closely matched to the silicone oil used for objective immersion. Images were acquired at ambient conditions, with the room climatised at 21 °C.

\blocktitle{Imaging of mouse embryonic stem cells}
Mouse embryonic stem cells expressing MCP-GFP and carrying MS2 loops inserted at the endogenous \textit{Sox2} locus were generated in-house. Cells were thawed and expanded for two passages prior to imaging to ensure recovery and stable growth. Cells were seeded onto Ibidi 35~mm glass-bottom dishes (1.5H coverslip thickness) coated with poly-L-ornithine and laminin to promote adhesion. Surfaces were incubated with 0.01\% poly-L-ornithine (Sigma-Aldrich P4957) overnight at 4 °C, washed with PBS, and then coated with laminin (L2020-1MG) diluted 1:200 in PBS  (5  $\mu$g/mL\ ;  1 mL per well, 5 µg total) for 2 h at 37 °C prior to cell seeding. 
Seeded dishes were incubated for 24~hours to allow the formation of layered colonies. Immediately before imaging, the growth medium was exchanged for an optically transparent imaging medium. During imaging the sample was kept at the optimal cell-growth conditions, i.e. 37 °C, 95\% humidity and 5\% CO2.

\blocktitle{Imaging of stem-cell–derived organoids}
Gastruloids were generated from the same MS2-labeled mESC line described above. Gastruloids were imaged at 120~hours after aggregation and mounted in Ibidi 35 mm glass-bottom dishes similarly coated with poly-L-ornithine .  Approximately 20 gastruloids were collected and embedded in 3D Life Dextran-PEG hydrogel FG as previously described \cite{Pineau2025}. A total volume of 60 $\mu$L of hydrogel at a final concentration of 0.8 mM was used to limit bulk motion while allowing continued growth. The hydrogel containing the gastruloids was deposited within a 0.50 mm-thick adhesive ring (Delta Microscopies, slide wells D70366-13), covered with a membrane (Isopore filter, 5.0 µm pore size, 13 mm diameter; Merck, TMTP01300), and secured with a second 0.25 mm-thick adhesive ring (Delta Microscopies, slide wells D70366-12). Dishes were centrifuged at 100g for 2 min and incubated for 13 min to allow gel polymerization, after which 2 mL of N2B27 medium was added on top. Analogously to what described for stem cells, the sample was kept at the optimal cell-growth conditions.


\section*{Acknowledgements}
We thank A.G. York, A. Millett-Sikking, J. Wong-Ng, G. Moneron, and the community of Optics Developers at Institut Pasteur. We acknowledge the use of AI-based writing tools for language enhancement. This work was supported by Institut Pasteur, Centre National de la Recherche Scientifique, grants from Région Ile-de-France (DIM ELICIT), the French National Research Agency (ANR-20-CE12-0028’ChroDynE’ and ANR-23-CE13-0021’GastruCyp’ and ANR-10 LABX-73’Revive’), and by funding from the European Research Council (ERC-2023-SyG, Dynatrans, 101118866). LF is also supported by CFM Foundation for Research.

\section*{Author contributions}
AS, MC, and TG conceptualized the work. AS and MC built microscope hardware and developed software. AS, MC, LF, and TG designed experiments and developed experimental protocols. AS, MC, and LF performed experiments. AS, LF, DK performed computational image analysis. CC generated the fluorescent reporter cell lines. AS, MC, and TG wrote the manuscript. MC and TG supervised the work. TG secured funding.

\section*{Competing interests}
The authors declare no competing interests.


\bibliographystyle{ieeetr}
\bibliography{bibliography}


\clearpage
\onecolumn
\begin{center}
  {\LARGE Supplementary Information}\\[1em]
  {\large An integrated quantitative single-objective light-sheet microscope\\ for subcellular dynamics in embryos and cultured multicellular systems}
\end{center}
\vspace{1em}

\section*{S1. Sampling constraints, image geometry, and 3D reconstruction}
\label{sec:sup_sampling}

The oblique illumination geometry of OPM introduces unique constraints on spatial sampling, image formation, and numerical reconstruction. Building on the coordinate transformation introduced in Sec.~\ref{sec:setup}, we describe how camera-recorded data in the tilted $(x',y',z')$ space are related to the laboratory frame $(x,y,z)$ (Fig.~\ref{fig:setup}), define sampling requirements for faithful reconstruction, and detail the deskewing and affine transformations applied during analysis. Galvo scanning and descanning generate a series of laterally undistorted oblique $(x',y')$ planes that are displaced along $z'$, producing a skewed volume that must be corrected by an affine deskewing transformation before further analysis.

\blocktitle{Sampling requirements and Nyquist criteria.}
The rotation of the image plane imposes anisotropic sampling that depends jointly on the camera pixel pitch, system magnification, and illumination tilt. For an effective pixel size $s_{\mathrm{cam}}$ on the camera and total magnification $M_{\mathrm{eff}}$, the in-plane sampling in the oblique frame is:
\[
s_{x'} = \frac{s_{\mathrm{cam}}}{M_{\mathrm{eff}}}, \qquad
s_{y'} = \frac{s_{\mathrm{cam}}}{M_{\mathrm{eff}}}.
\]
After rotation back to the laboratory frame, the effective sampling becomes:
\[
s_{y} = s_{y'} \cos\theta, \qquad
s_{z} = s_{y'} \sin\theta.
\]
Thus, for a given tilt angle $\theta$, the $y$ and $z$ sampling intervals differ by a factor of $\tan\theta$.

\blocktitle{Galvo scanning and uniform slice spacing.}
To sample the volume uniformly, the galvanometric mirror is driven with a calibrated stair-step waveform that ensures linear scanning in $z'$. Any nonlinearity in the galvo response would introduce compression or stretching of the reconstructed volume along $z$. For a galvo scan amplitude $A$ and optical geometry defined by the conjugate pupil relationships, the slice spacing in the oblique frame is:
\[
\Delta z' = \frac{\alpha A}{N_{\mathrm{slices}}},
\]
where $\alpha$ is the angle-to-lateral-displacement conversion factor. The laboratory-frame slice spacing is then:
\[
\Delta z = \Delta z' \cos\theta.
\]

\blocktitle{Deskewing and affine transformation.}
Raw OPM images appear sheared because each oblique slice is shifted laterally by an amount proportional to its axial position. Reconstruction consists of deskewing (undoing lateral shear) followed by an affine rotation mapping the tilted $y'–z'$ plane back to the laboratory frame. Because both operations are linear, the transformation can be represented as a $3 \times 3$ affine matrix and implemented efficiently during analysis.

\blocktitle{Field-of-view and clipping constraints.}
For sufficiently large scanning amplitudes or wide fields of view, the remote-refocusing pupil can partially occlude the tilted imaging plane, producing PSF broadening and reduced signal. This effect is a direct geometric consequence of remote refocusing and sets a fundamental limit on the usable scan range. We quantify this limit empirically by scanning fluorescent beads across the volume and identifying clipping regions where the PSF broadens or the signal drops.

\blocktitle{Mechanical layout and sample geometry.}
To support homogeneous illumination and quantitative imaging across the usable field of view, the excitation beam is pre-shaped using a Powell lens and a series of cylindrical lenses. A resonant mirror is incorporated to suppress stripe artefacts, and an adjustable slit placed in a plane conjugate to the illumination pupil allows tuning of the effective light-sheet numerical aperture. Sample positioning is achieved using a piezo stage, with optional coarse positioning, to place the region of interest reproducibly within the imaging volume. Fluorescence detection is performed using dual-camera imaging, in which emitted light is spectrally separated by a dichroic beamsplitter depending on the imaging condition, with each camera equipped with its own emission filter set. The primary sCMOS camera records oblique-plane fluorescence images for quantitative analysis, while a secondary widefield camera provides sample context and facilitates routine alignment and targeting. In addition, a dedicated auxiliary camera positioned upstream of the main emission dichroic is used exclusively for back-reflection–based autofocus and drift compensation (Sec.~\ref{sec:stabilization}), without introducing additional optics into the fluorescence detection path.

\section*{S2. Photon budget, transmission, and camera noise}
\label{sec:sup_photon}


Accurate quantitative imaging requires not only the optical characterization described in Secs.~\ref{sec:setup}–\ref{sec:psf} but also a detailed understanding of how efficiently photons are transmitted through the detection path and converted into measurable electronic signals. Because many biological measurements operate near the photon-limited regime—such as transcription-site intensity extraction and faint locus localization—the overall photon budget sets the achievable signal-to-noise ratio.

\blocktitle{Optical transmission of the detection path.}
The fraction of detected photons is determined jointly by the optical transmission of the relay optics and the detector QE. To quantify this contribution, we measured the throughput from the primary objective to the camera by recording a calibrated fluorescence signal before and after the remote-refocusing module and emission filters. Across 488, 561, and 594\,nm excitation, the total transmission of the detection chain ranged from 40–55\%, consistent with expected losses at each interface. This transmission factor enters directly into downstream estimates of photon budgets and localization uncertainties.

\blocktitle{Photon-transfer measurement of system gain.}
To convert measured pixel values into photon counts, we performed a photon-transfer calibration by acquiring uniform-field images at increasing brightness and analyzing the mean–variance relationship. The resulting photon-transfer curve shows the expected linear dependence in the shot-noise regime, with slope and intercept yielding the system gain $g$ and read noise $\sigma_{\mathrm{read}}$, respectively. Photon-transfer analysis is a widely used method in detector characterization \cite{Janesick2001}.

Across all settings used in OPM acquisition, we measured $g = 0.43 \pm 0.01~\mathrm{e^{-}/ADU}$ and $\sigma_{\mathrm{read}} = 1.2$--$1.5~\mathrm{e^{-}}$.

\blocktitle{Read noise, fixed-pattern noise, and hot pixels.}
The electronic noise floor of the detector was quantified using dark-frame sequences acquired at the same frame rates used during OPM imaging. The temporal distribution of pixel values is Gaussian with standard deviation matching the PTC-derived $\sigma_{\mathrm{read}}$, indicating stable and stationary read noise. Spatially correlated noise components—fixed-pattern noise (FPN)—were extracted by subtracting long-term pixel averages. FPN is a known feature of sCMOS detectors and is typically corrected through offset/dark-frame subtraction \cite{Waters2009}.

Hot pixels were identified as those exhibiting persistent offsets exceeding five times the temporal noise level. These pixels accounted for less than 0.02\% of the sensor array and were masked or interpolated during preprocessing. After correction, hot pixels and FPN contributed negligibly to the noise budget.

\blocktitle{Detector quantum efficiency and linearity.}
The detector’s quantum efficiency (QE) and linearity were characterized using uniform-field exposures across a range of intensities and exposure times. The mean signal remained linear over more than 95\% of the full-well capacity, with deviations only near saturation. Combining QE with optical transmission measurements yields the total photon-detection efficiency used in quantitative intensity interpretation.

\section*{S3. Image analysis: segmentation, spot detection, and tracking}
\label{sec:sup_analysis}


Quantitative biological imaging requires robust segmentation, accurate detection of fluorescent puncta, and reliable tracking across space and time. Because OPM data consist of large volumes with spatially varying SNR and anisotropic resolution (Sec.~\ref{sec:psf}), image analysis must incorporate both classical computer-vision algorithms and modern neural-network-based approaches. The procedures summarized here draw on established practices in bioimage analysis \cite{Ronneberger2015, Stringer2021, Sage2019} and are adapted to the specific geometry of OPM (Sec.~\ref{sec:setup}).

\blocktitle{3D nuclei segmentation.}
Nuclei were segmented from reconstructed OPM volumes using a hybrid strategy combining thresholding-based initialization with a network-refined segmentation stage. Initial masks were obtained by smoothing, intensity normalization, and adaptive thresholding, followed by connected-component labeling. These coarse masks were then refined using a U-Net or Cellpose-type architecture trained on manually annotated sub-volumes to accommodate the OPM-specific anisotropy and depth-dependent contrast.

\blocktitle{Single-particle detection.}
Fluorescent puncta corresponding to transcription sites, tagged loci, or single mRNAs were detected using a combination of bandpass filtering and local maxima extraction. Candidate particle positions were then filtered by intensity, shape, and SNR criteria informed by the detector noise model (Supplementary Sec.~S2), following established practices in single-molecule imaging.

\blocktitle{Sub-pixel localization by Gaussian fitting.}
For each detected spot, sub-pixel localization was performed by fitting a three-dimensional Gaussian to a small region surrounding the candidate maximum. Because the OPM PSF is anisotropic (Sec.~\ref{sec:psf}), fitting is performed with independent widths along each axis.

\blocktitle{Trajectory reconstruction.}
Particle trajectories were reconstructed across frames using a global assignment algorithm that minimizes the total cost of associating detections between successive time points \cite{Jaqaman2008}. The cost function incorporates spatial proximity, expected motion models, and localization uncertainties estimated from the Gaussian fits.

\section*{S4. Extended biological applications: single-molecule detection and tracking}
\label{sec:sup_bio}

This section provides extended application examples that are not central to the main-text transcription quantification claims, including single-molecule detection and short-duration tracking analyses in embryos and mammalian systems.

\blocktitle{Single mRNA detection and tracking in \textit{Drosophila} embryos.}
At later stages or in constructs with lower expression, individual cytoplasmic mRNAs appear as isolated fluorescent puncta. Detection uses the bandpass filtering + local-maxima pipeline (Supplementary Sec.~S3), with SNR thresholds informed by the detector model (Supplementary Sec.~S2). Detected single mRNAs can be tracked across frames using global assignment (Supplementary Sec.~S3), enabling mean-squared displacement (MSD) analysis and compartment-specific comparisons.

\blocktitle{Single-molecule detection in mESCs and gastruloids.}
In mESCs and gastruloids, the extent to which absolute single-molecule quantification is achievable depends on photobleaching, background structure, and available calibration strategies. In systems with currently available reporters (e.g.\ SOX2), photobleaching is measurable and approximately linear over acquisition windows, but defining an absolute cytoplasmic unit remains an open question. Potential calibration approaches include population-averaged comparisons and future incorporation of additional tagged genes or cytoplasmic standards.

\section*{S5. Mechanical implementation and sample positioning}
\label{sec:sup_mechanics}

The optical layout described in the main text is implemented on a compact inverted microscope platform designed to maximize mechanical stability while preserving flexibility in sample handling. The primary objective (OBJ1) serves as the fixed optical reference for the entire system and is rigidly mounted on a dedicated breadboard featuring a central aperture. This mounting geometry ensures stable alignment of OBJ1 and all conjugate optical planes, while avoiding moving optical elements during routine focusing or acquisition.

Sample positioning is achieved by translating the specimen rather than the objective. The sample holder is mounted on a motorized stage providing coarse positioning along the $x$, $y$, and $z$ axes, combined with a piezoelectric actuator for fine axial control. This configuration allows precise and reproducible placement of the region of interest within the imaging volume while maintaining a fixed optical reference frame defined by OBJ1. During focusing and volumetric acquisition, the position of the primary objective and associated relay optics remains unchanged, ensuring consistent optical alignment across experiments.

The sample stage supports a compact environmental enclosure that accommodates live imaging under controlled temperature and atmospheric conditions. This design enables imaging of diverse biological specimens, including \textit{Drosophila} embryos, adherent mammalian cells, and three-dimensional aggregates, while maintaining compatibility with standard coverslip-based mounting formats. The mechanical stability of the platform minimizes drift during long time-lapse experiments and complements the optical autofocus and stabilization procedures described in Supplementary Sec.~S\ref{sec:stabilization}.


\section*{S6 Comparison to existing OPM and single-objective light-sheet platforms}
\label{sec:comparison}

\begin{figure}[t]
  \centering
  \caption{\textbf{Comparison to existing single-objective light-sheet platforms.}
  (\textbf{A}) Summary of key specifications (NA, field of view, volume rate, quantitative characterization).
  (\textbf{B}) Performance comparison in terms of resolution, imaging volume, and volumetric acquisition rate.}
  \label{fig:comparison}
\end{figure}

Single-objective light-sheet microscopy has advanced rapidly over the past decade, with systems such as SCAPE2.0, eSPIM, SOPi, and DaXi providing increasingly high NA, fast volume rates, and improved compatibility with standard sample geometries. These platforms share the central design principle of using a single primary objective for both illumination and detection, but differ substantially in optical layout, achievable NA, volumetric scan strategy, and quantitative characterization.
Here we compare our OPM implementation to these established systems using published specifications and performance benchmarks (Fig.~\ref{fig:comparison}), focusing specifically on quantitative metrics relevant to transcription imaging and nuclear-scale dynamics.

\blocktitle{Quantitative comparison of specifications.}
Figure~\ref{fig:comparison} summarizes key specifications of our OPM system alongside representative values reported for SCAPE2.0 \cite{Voleti2019}, eSPIM \cite{Sapoznik2020}, SOPi \cite{Kumar2018}, and DaXi \cite{Yang2022}. These include numerical aperture, effective resolution, field of view, imaging volume, volumetric acquisition rate, and degree of quantitative optical characterization.

\blocktitle{Resolution, imaging volume, and volumetric rate.}
Our OPM implementation occupies a regime characterized by high numerical aperture, moderate-to-high volumetric acquisition rates, a large usable field of view, and quantitative characterization of PSF, sampling geometry, and timing synchronization (Secs.~\ref{sec:setup}–\ref{sec:software}). This combination positions our system within the landscape of single-objective light-sheet microscopy and specifically supports quantitative biological applications—such as MS2 transcription imaging—that require both precision and robustness while remaining compatible with standard sample geometries.

\end{document}